\documentclass[runningheads]{llncs}
\usepackage[T1]{fontenc}
\usepackage{graphicx}
\usepackage{rotating}
\usepackage{cite}
\usepackage{amsmath}
\usepackage{amssymb}
\usepackage{booktabs}
\usepackage{array}
\usepackage{multirow} 

\begin{document}
%
\title{QMC-Net: Data-Aware Quantum Representations for Remote Sensing Image Classification}
%
%
\author{Md Aminur Hossain\inst{1}\orcidID{0009-0003-6357-7480} \and
Ayush V. Patel\inst{4}\orcidID{0000-0002-8921-7091} \and
Biplab Banerjee\inst{2, 3}\orcidID{0000-0001-8371-8138}}
\authorrunning{MA. Hossain et al.}
%
\institute{Space Applications Centre, Indian Space Research Organisation, Ahmedabad, India \and
Centre of Studies in Resources Engineering, Indian Institute of Technology, Bombay, India\\ \and
Centre for Machine Intelligence and Data Science, Indian Institute of Technology, Bombay, India\\ \and
School of Technology, Pandit Deendayal Energy University, Gandhinagar, India\\
\email{\{md.aminurhossain,ayu020503,getbiplab\}@gmail.com}}
\maketitle              
\begin{abstract}
Hybrid quantum–classical models offer a promising route for learning from complex data; however, their application to multi-band remote sensing imagery often relies on generic, data-agnostic quantum circuits that fail to account for channel-specific statistical variability. In this work, we propose a data-driven framework that maps band-level statistics such as Shannon Entropy, Variance, Spectral Flatness, and Edge Density to the hyperparameters of customized quantum circuits. Building on this framework, we introduce QMC-Net, a hybrid architecture that processes six data channels using band-specific quantum circuits, enabling adaptive quantum feature encoding and transformation across channels. Experiments on the EuroSAT and SAT-6 datasets demonstrate that QMC-Net achieves accuracies of 93.80\% and 99.34\%, respectively, while a residual-enhanced variant further improves performance to 94.69\% and 99.39\%. These results consistently outperform strong classical baselines and monolithic hybrid quantum models, highlighting the effectiveness of data-aware quantum circuit design under NISQ constraints.

\end{abstract}
%
%
\section{Introduction}
\label{sec:introduction}
The growing availability of multi-band satellite data has led to the widespread adoption of deep learning approaches, particularly Convolutional Neural Networks (CNNs), in Earth observation (EO) \cite{rodriguez2020earth}. These models have shown considerable promise, achieving strong benchmark performance in land-use classification on datasets such as EuroSAT \cite{helber2019eurosat}. However, the underlying architectural paradigm is fundamentally monolithic: all data channels are subjected to the same set of convolution operations despite their potentially vastly different statistical and physical characteristics. As the community begins to assess the possibilities of Quantum Machine Learning (QML) \cite{schuld2015introduction}, this agnostic data-processing philosophy is often inherited by hybrid quantum–classical models, which may limit their ability to effectively exploit quantum representations.

The advent of the Noisy Intermediate-Scale Quantum (NISQ) paradigm has sparked renewed interest in hybrid quantum–classical models, in which Parameterized Quantum Circuits (PQCs) are employed as specialized co-processors within classical deep learning pipelines \cite{deshpande2024dynamic}. In this setting, quantum circuits are not expected to replace large classical models, but rather to provide compact, expressive transformations under strict hardware constraints. Early studies have demonstrated that such hybrid architectures can be applied to remote sensing tasks, achieving performance comparable to classical counterparts while using PQCs as high-dimensional feature extractors \cite{benedetti2019parameterized}. However, the NISQ regime fundamentally limits circuit depth, qubit count, and coherence time, making it impractical to scale quantum components in the same manner as classical deep networks. As quantum models grow larger and deeper, noise accumulation and trainability issues can suppress genuine quantum effects, reducing the circuit to behavior that is effectively classical.

Nevertheless, a recurring limitation in the existing literature is the use of a uniform quantum circuit design, where a single generic parameterized quantum circuit (PQC) is applied across all input channels, ranging from raw spectral bands to complex biophysical indices. This approach implicitly assumes that all data channels present comparable computational demands when processed by a quantum circuit. In practice, this assumption leads to two key inefficiencies: circuits that are too simple may underfit complex channels and create information bottlenecks, while overly expressive circuits consume limited quantum resources such as qubits and coherence time when applied to simpler data.

This work addresses this limitation by moving beyond data-agnostic quantum circuit design. We hypothesize that tailoring circuit architectures to the statistical complexity of individual data channels can improve classification performance while using quantum resources more efficiently. To this end, we introduce a principled, data-driven framework that links classical data statistics, namely Shannon Entropy, Variance, Spectral Flatness, and Edge Density, to quantum circuit hyperparameters including circuit width, depth, entanglement strategy, and parameterization density. The main contributions of this work are as follows:
\begin{enumerate}
    \item We propose a formal framework for the data-driven design of customized quantum circuits, establishing a direct link between classical data statistics and quantum architectural decisions.
    \item We introduce QMC-Net, a hybrid quantum–classical architecture that instantiates this framework using band-specific quantum circuits to process a six-channel input composed of RGB bands and three derived channels.
\end{enumerate}

\section{Related Work}
\label{sec:related_works}

Our work lies at the intersection of three research areas: deep learning for remote sensing, quantum machine learning for Earth Observation (EO), and data-dependent quantum representations.

Modern EO is characterized by large-scale, high-volume data streams. EO is a field of diverse satellite missions that produce large, continuous streams of data. This has led to a shift towards advanced Artificial Intelligence (AI) techniques \cite{ESA2021ai}. In particular, Deep Learning has been demonstrated as a state-of-the-art approach for many remote sensing applications using Convolutional Neural Networks (CNNs) \cite{o2015introduction}, such as land-use and land-cover (LULC) identification and classification \cite{lou2025land} on benchmark datasets (e.g. EuroSAT \cite{ helber2019eurosat}, Aerial Image Dataset \cite{xia2017aid}, etc). These models are highly effective at learning hierarchical spatial features directly from pixel-level inputs \cite{kattenborn2021review}. However, standard CNN architectures typically apply identical convolutional operations across all input channels, an assumption that may be sub-optimal when dealing with the heterogeneous statistical properties inherent in multi-band satellite imagery.

Motivated by the exploration of next-generation computing paradigms for remote sensing, quantum computing has recently emerged as a promising direction. Within the current Noisy Intermediate-Scale Quantum (NISQ) regime, most practical quantum machine learning approaches adopt hybrid quantum–classical algorithms \cite{bharti2021noisy}, where parameterized quantum circuits (PQCs) are integrated into classical optimization pipelines \cite{tychola2023quantum}. Early studies have demonstrated the feasibility of such hybrid models for remote sensing applications, particularly through quantum convolutional neural networks (QCNNs) \cite{rajesh2021quantum, fan2024land}, which employ PQCs as feature extractors within classical CNN frameworks. This line of research originates from the concept of quanvolutional neural networks \cite{henderson2019quanvolutional} and includes early proof-of-concept studies showing that quantum circuit-based models can effectively process multi-spectral satellite data \cite{gawron2020multi}. Subsequent work has explored alternative PQC architectures and has shown that entangled circuits generally outperform non-entangled counterparts on datasets such as EuroSAT \cite{sebastianelli2021circuit}.

Despite this progress, most existing hybrid quantum–classical models for remote sensing adopt a uniform circuit design, applying the same PQC architecture to all input channels regardless of their underlying statistical characteristics. In practice, remote sensing data channels exhibit substantial diversity; for example, raw RGB bands, biophysical indices such as Normalized Difference Vegetation Index (NDVI), and textural entropy maps represent fundamentally different data distributions with distinct spatial and statistical properties. Prior work in the broader quantum machine learning literature has established that the expressivity, trainability, and performance of quantum neural networks are strongly influenced by both circuit architecture and input data characteristics \cite{huang2021power}. However, the specific problem of designing data-dependent quantum architectures for multi-modal and statistically heterogeneous remote sensing data remains largely unexplored. This gap motivates data-driven quantum circuit designs tailored to the statistical complexity of individual data channels.

\section{Proposed Methodology}
The central hypothesis underlying this work is that tailoring quantum circuit architectures to the statistical characteristics of individual data channels can yield improved model performance compared to a monolithic design. This section describes the construction of a multi-modal dataset from satellite imagery and outlines the feature engineering process that supports the design of band-specific quantum circuits.

\begin{figure*}[t]
    \centering
    \includegraphics[width=\textwidth]{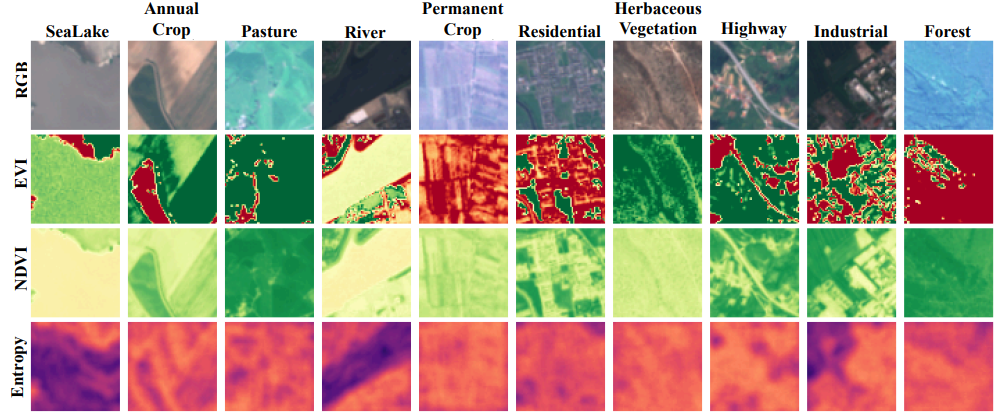}
    \caption{
        Visualization of the six-channel data (RGB, EVI, NDVI, \& Entropy) derived from the EuroSAT dataset for representative samples of all 10 classes.
    }
    \label{fig:band-visualization}
\end{figure*}

\subsection{Datasets and Feature Engineering}
We utilize the EuroSAT and DeepSAT (SAT-6) \cite{basu2015deepsat} datasets, which are widely used benchmarks for land-use and land-cover classification. EuroSAT consists of 27,000 labeled $64 \times 64$ Sentinel-2 images spanning 10 classes and provides 13 spectral bands, from which we select the RGB and near-infrared (NIR) bands. DeepSAT (SAT-6) contains 405,000 samples derived from the National Agricultural Imagery Program (NAIP) across six classes, with images of size $28 \times 28$ and four spectral bands. From these selected bands, we construct a six-channel input tensor that captures complementary spectral, biophysical, and textural information, as described below:

\begin{itemize}
    \item \textbf{Visible Spectrum (Red, Green, Blue):} The visible spectrum is represented by the Red, Green, and Blue channels, corresponding to Sentinel-2 Bands 4, 3, and 2, respectively for EuroSAT. These channels provide the basis for standard color imagery and human visual interpretation.
    \item \textbf{Biophysical Indices (NDVI, EVI):} We engineered two standard remote sensing indices to summarize vegetation health and density. NDVI is a measure of vegetation vigor, created using the Red (B4) and Near-Infrared (B8) bands. The Enhanced Vegetation Index (EVI) is an improved index that alleviates saturation over dense canopies while also reducing atmospheric effects using the Blue band (B2). These indices are defined as:
    \begin{align}
        \text{NDVI} &= \frac{\text{NIR} - \text{Red}}{\text{NIR} + \text{Red} + \epsilon}, & 
        \text{EVI} &= 2.5 \times \frac{\text{NIR} - \text{Red}}{\text{NIR} + 6 \times \text{Red} - 7.5 \times \text{Blue} + 1}
    \end{align}

    \item \textbf{Textural Information (Entropy):} To capture local spatial heterogeneity, we generate a textural entropy map $x$ by applying a rank-based entropy filter with a 5-pixel radius disk to a grayscale representation of the RGB image. Grayscale conversion is used to capture structural texture independent of spectral color. The entropy is computed as:
    \begin{multline}
        \text{Entropy}(x) = \text{Normalize}(\text{RankEntropy}(\text{Gray}(R,G,B),
        \text{disk}(r=5)))
    \end{multline}
    The resultant values are normalized to the range $[0,1]$.  Together, these six engineered channels form the multi-modal input to our model, as illustrated in Fig.~\ref{fig:band-visualization} for EuroSAT dataset.
\end{itemize}

For data splits, EuroSAT was partitioned using a 70:15:15 ratio, yielding 18,900 training images and 4,050 images each for validation and testing. For SAT-6, a stratified 10\% subsample was used, producing 32,400 training and 8,100 testing images, with a further 20\% of the training set reserved for validation.

\subsection{Metric-Driven Quantum Circuit Design Framework}
\label{sec:framework}
Designing an effective quantum machine learning (QML) model requires a circuit architecture that is sufficiently expressive to capture the underlying structure of the data, while remaining sufficiently constrained to ensure trainability within the limitations of the Noisy Intermediate-Scale Quantum (NISQ) regime. In practice, NISQ hardware imposes strict limits on circuit depth, qubit count, and coherence time, and overly complex circuits are prone to issues such as barren plateaus and noise-induced degradation. In this subsection, we present a principled framework that systematically maps the statistical properties of classical input data to the architectural hyperparameters of a quantum circuit, enabling a balanced trade-off between expressivity and feasibility.

\subsubsection{Data Complexity Metrics:}
\label{sec:metrics}
For each image band, pixel values are normalized to the range $[0,255]$ to enable consistent computation of statistical metrics across bands with different native dynamic ranges. Let an image band be represented by a matrix of pixel intensities $I$. We compute four data complexity metrics defined below:

\begin{itemize}
    \item \textbf{Shannon Entropy ($\boldsymbol{H}$):} A measure of information content computed from the normalized pixel intensity histogram. Let $p(i)$ denote the probability of pixels with intensity $i \in \{0,\ldots,255\}$. Entropy is defined as
    \begin{equation}
        H(I) = -\sum_{i=0}^{255} p(i)\log_{2}p(i),
    \end{equation}
    where a small constant is added to $p(i)$ for numerical stability. Higher entropy indicates greater data complexity.
    
    \item \textbf{Variance ($\boldsymbol{\sigma^2}$):} Variance quantifies the dispersion of pixel intensities around the mean $\mu$ and is given by
    \begin{equation}
        \sigma^2(I) = \frac{1}{N} \sum_{j=1}^{N} (x_j - \mu)^2,
    \end{equation}
    where $x_j$ are pixel values and $N$ is the total number of pixels. Higher variance reflects a broader intensity range.
    
    \item \textbf{Spectral Flatness ($\boldsymbol{F}$):} Spectral flatness measures the uniformity of the intensity distribution and is defined as the ratio of the geometric mean to the arithmetic mean of pixel values $x_j$,
    \begin{equation}
        F(I) = \frac{\left(\prod_{j=1}^{N} x_j\right)^{1/N}}{\frac{1}{N}\sum_{j=1}^{N} x_j},
    \end{equation}
    with a small constant added for numerical stability. Values near 1 indicate uniform distributions, while values near 0 indicate peaked distributions.
    
    \item \textbf{Edge Density ($\boldsymbol{ED}$):} Edge density captures local spatial correlations by computing the proportion of pixels classified as edges. Using the Sobel operator to compute gradient magnitude $G(p)$, edge density is defined as
    \begin{equation}
        ED(I) = \frac{|\{p \in I : G(p) > \theta\}|}{N},
    \end{equation}
    where $N$ is the total number of pixels and $\theta=0.1$. Higher edge density indicates stronger spatial correlations.
\end{itemize}

\subsubsection{Justification of the Metric-to-Circuit Correspondences:}
\label{sec:justification}
At the core of our framework are four primary mappings that relate classical data metrics to architectural decisions in quantum circuit design. We provide a formal justification for each mapping based on principles of information theory, quantum computing, and function approximation.

\begin{enumerate}
    \item \textbf{Flatness ($\boldsymbol{F}$) and Shannon Entropy ($\boldsymbol{H}$) to Circuit Width ($\boldsymbol{Q}$):}
    Shannon entropy $H$ provides a natural measure of the information content that must be represented by a quantum circuit with $Q$ qubits. From information theory, the number of distinguishable classical states can be approximated as $N \approx 2^H$. To avoid information loss, the Hilbert space dimension $2^Q$ must be sufficient to represent these states, leading to
    \begin{equation}
    2^Q \gtrsim N \;\Rightarrow\; Q \gtrsim H
    \end{equation}
    This constraint mitigates information bottlenecks during quantum data encoding \cite{schuld2021effect}. Spectral flatness $F$ modulates this requirement: high flatness implies nearly equiprobable states, making $Q \gtrsim H$ a strict constraint, whereas low flatness indicates sparse distributions, allowing it to be treated as a soft guideline and enabling more efficient resource allocation.
    
    \item \textbf{Shannon Entropy ($\boldsymbol{H}$) to Circuit Depth ($\boldsymbol{D}$):} 
    A variational quantum circuit acts as a function approximator whose expressive capacity is closely tied to its depth $D$. The circuit unitary is a composition of $D$ layers $U(\boldsymbol{\theta}) = U_D(\boldsymbol{\theta}_D) \cdots U_1(\boldsymbol{\theta}_1)$ with expressivity increasing as depth grows \cite{sim2019expressibility}. High-entropy data contains complex structure that requires a high-capacity model to avoid underfitting. A greater entropy $H$ means greater model capacity is necessary, which can be directly provided by increasing depth $D$. However, excessive depth increases the risk of barren plateaus, where gradients vanish exponentially and training becomes impractical \cite{mcclean2018barren}.

    \item \textbf{Edge Density ($\boldsymbol{ED}$) to Entanglement Strategy ($\boldsymbol{E}$):}
    Edge density reflects the strength of local statistical dependencies in image data, with edges indicating strong correlations between neighboring pixels. To model such dependencies, a quantum circuit must produce correlated measurement outcomes across qubits. The correlation between measurements on qubits $A$ and $B$ is given by the two-point correlation function $C(A,B) = \langle \sigma_z^A \otimes \sigma_z^B \rangle$. Circuits without entangling gates generate separable states $|\psi_{\text{sep}}\rangle = |\psi_A\rangle \otimes |\psi_B\rangle$, for which correlations factorize as
    \begin{equation}
        C_{\text{sep}}(A,B) = \langle \sigma_z^A \rangle \langle \sigma_z^B \rangle,
    \end{equation}
    and are therefore unable to represent the covariance present in complex edge patterns. Capturing high edge density thus requires entangled states for which this factorization no longer holds. Leveraging entanglement to represent classical correlation structures is well established in quantum machine learning \cite{zoufal2019quantum}, motivating denser entanglement strategies for channels with higher edge density.

    \item \textbf{Variance ($\boldsymbol{\sigma^2}$) to Parameterization Density ($\boldsymbol{P}$):} A variational circuit defines a function $f_{\boldsymbol{\theta}}(x)$ and the capability to approximate more complex target functions depends on the expressivity of its parameterized gates. Variational circuits can be understood as a general class of Fourier series, in which the richness of the function space is determined by the set of gates used \cite{schuld2021effect}.  For data with high variance, a more complex, non-linear mapping is required for accurate approximation. Circuits composed of simple single-parameter gates (e.g., $R_Y(\theta)$) are limited in the Fourier features they can represent. In contrast, three-parameter gates form a universal representation for single-qubit unitaries,
        \begin{equation}
        U(\alpha, \beta, \gamma) = R_Z(\alpha)R_Y(\beta)R_Z(\gamma)
    \end{equation}
    providing a richer basis and improved expressivity. This property is central to constructing universal quantum function approximators \cite{perez2020data}.
\end{enumerate}

The above principles are summarized in the design rubric shown in Table~\ref{tab:design_rubric}, which translates quantitative data analysis into actionable architectural decisions under practical constraints, including a maximum circuit width of eight qubits. This rubric guides the design of four custom circuits for EuroSAT and two for SAT-6 based on the classical properties in Table~\ref{tab:classical_stats}, demonstrating the practical utility of the proposed framework across all six input bands. These mappings serve as principled design heuristics rather than strict optimality guarantees.

\begin{table}[h!]
    \caption{The Metric-Driven Design Rubric. This table links each quantum hyperparameter to its primary classical data driver and the corresponding design heuristic.}
    \label{tab:design_rubric}
    \centering
    \scriptsize
    \renewcommand{\arraystretch}{1.4}
    \setlength{\tabcolsep}{6pt}
    \begin{tabular}{|l|l|>{\raggedright\arraybackslash}p{3.5cm}|}
        \hline
        \textbf{Hyperparameter} & \textbf{Driving Metric(s)} & \textbf{Design Heuristic} \\
        \hline
        \textbf{Qubits ($Q$)} & Entropy ($H$), Flatness ($F$) & $Q \gtrsim H$, strict if $F$ is high \\
        \hline
        \textbf{Depth ($D$)} & Entropy ($H$) & Low $\to$ 1-2L; Med $\to$ 2-4L; High $\to$ 4+L$^*$ \\
        \hline
        \textbf{Entanglement ($E$)} & Edge Density ($ED$) & Low $\to$ Linear; Med $\to$ Ring; High $\to$ Multi-scale \\
        \hline
        \textbf{Parameters ($P$)} & Variance ($\sigma^2$) & Low $\to$ 1/Q; Med $\to$ 3/Q; High $\to$ 3/Q + Re-upload \\
        \hline
        \multicolumn{3}{|l|}{\scriptsize{$^*$Increased depth is applied cautiously to mitigate barren plateaus.}} \\
        \hline
    \end{tabular}
\end{table}

\begin{table}[t]
\centering
\scriptsize
\renewcommand{\arraystretch}{1.3}
\setlength{\tabcolsep}{4pt}

\caption{Classical metrics for EuroSAT and SAT-6 datasets}
\label{tab:classical_stats}

\begin{tabular}{|l|l|c|c|c|c|}
\hline
\textbf{Dataset} & \textbf{Band} & $\boldsymbol{H}$ & $\boldsymbol{\sigma^2}$ & $\boldsymbol{F}$ & $\boldsymbol{ED}$ \\
\hline
\multirow{6}{*}{EuroSAT}
& Red     & 6.4017 & 4015.0316 & 0.2109 & 0.5775 \\
& Green   & 6.6672 & 4022.5922 & 0.2332 & 0.5898 \\
& Blue    & 6.7971 & 4019.2067 & 0.2401 & 0.5598 \\
& NDVI    & 7.0133 & 2911.9893 & 0.8526 & 0.3098 \\
& EVI     & 2.9599 & 6512.0394 & 0.2438 & 0.3368 \\
& Entropy & 7.1627 & 1888.4368 & 0.9420 & 0.1369 \\
\hline
\multirow{6}{*}{SAT-6}
& Red     & 7.2234 & 2111.5678 & 0.8547 & 0.2305 \\
& Green   & 7.1720 & 1985.1396 & 0.8752 & 0.2882 \\
& Blue    & 7.2317 & 2020.6025 & 0.8880 & 0.1890 \\
& NDVI    & 7.0817 & 1675.5893 & 0.9004 & 0.1712 \\
& EVI     & 4.4453 & 694.8035  & 0.9063 & 0.0495 \\
& Entropy & 6.9983 & 1708.5580 & 0.9585 & 0.2407 \\
\hline
\end{tabular}

\end{table}

\subsection{Hybrid Quantum-Classical Model Architecture}
\begin{figure*}[h]
    \centering
    \includegraphics[width=\textwidth]{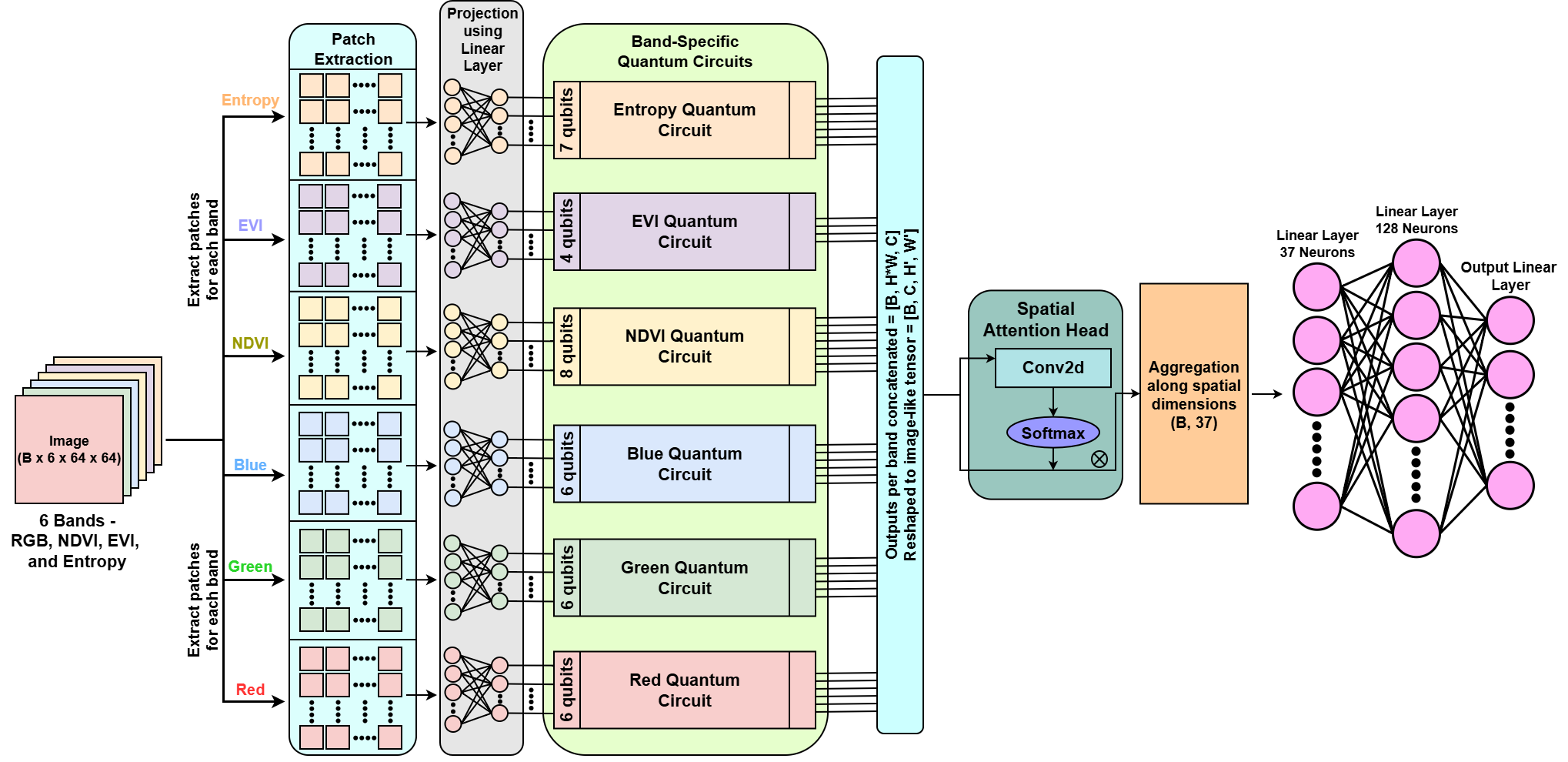} 
    \caption{QMC-Net Architecture: A hybrid quantum-classical deep learning framework for processing multi-band remote sensing imagery (RGB, EVI, NDVI, and Entropy).
    }
    \label{fig:overall_architecture}
\end{figure*}

Our model, Quantum Multi Channel Network (QMC-Net), is a hybrid quantum–classical framework in which distinct quantum circuits are used to analyze different data channels, including spectral bands and derived biophysical and textural features, from satellite imagery. The top-level illustration of the architecture is provided in Fig.~\ref{fig:overall_architecture}. The model is composed of two major sub-components: a Quantum Feature Encoder working to extract features from image patches utilizing band-specific circuits, and a classical Spatial Attention Head that performs the final classification.

\begin{figure}[t]
    \centering
    \includegraphics[width=\textwidth]{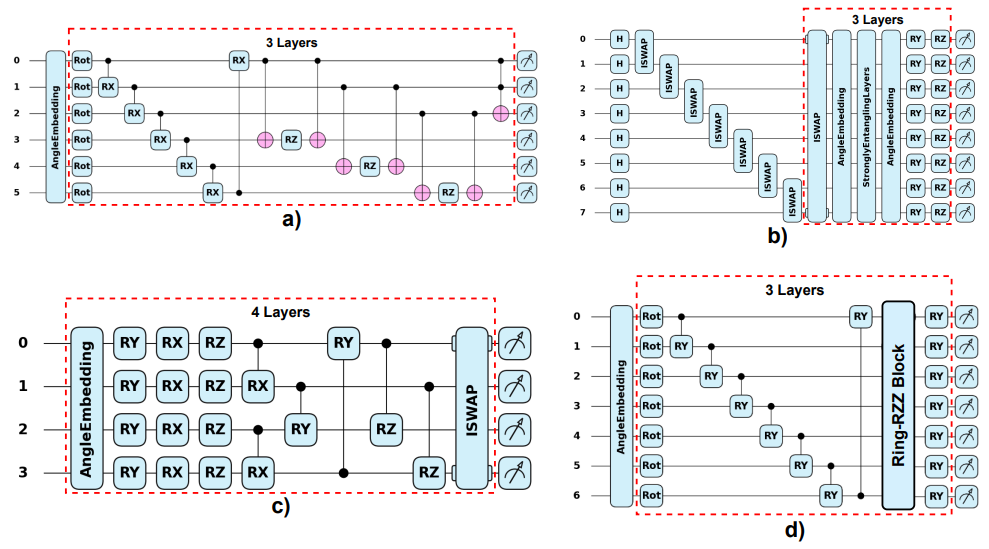}
    \caption{
        Band-specific Quantum Circuits designed for EuroSAT bands by following the design rubrics - a) RGB Circuit, b) NDVI Circuit, c) EVI Circuit, \& d) Entropy Circuit
    }
    \label{fig:quantum_circuits}
\end{figure}

\subsubsection{Quantum Feature Encoder:}
The goal of the encoder is to transform classical image data into quantum feature representations. This process consists of three sequential steps:
\begin{itemize}
    \item \textbf{Classical Pre-processing:} Each of the six input bands is partitioned into non-overlapping patches, reshaped into one-dimensional vectors, and passed through a band-specific trainable linear layer that projects them to a dimension matching the qubit count of the corresponding quantum circuit.
    \item \textbf{Band-Specific Quantum Circuits:} The projected classical vector from each band is provided as input to its associated quantum circuit, and the output of each circuit is obtained as the expectation value of the Pauli-Z operator $\langle \sigma_z \rangle$, due to its stability and interpretability measured on each qubit. While bands are processed independently, bands with statistically similar distributions share the same circuit architecture. In particular, the RGB bands of EuroSAT and the RGB, NDVI, and entropy bands of SAT-6 exhibit nearly indistinguishable statistical characteristics (Table~\ref{tab:classical_stats}), motivating the use of a common circuit design for these channels. The four quantum circuit architectures designed for the EuroSAT bands are described below:
    
    \begin{itemize}
        \item \textbf{RGB Bands Circuit:} The RGB bands exhibit high edge density ($ED \approx 0.58$), motivating a circuit with a rich entanglement strategy to model strong spatial dependencies. Accordingly, we employ a 6-qubit, 3-layer circuit, as shown in Fig.~\ref{fig:quantum_circuits}-a).
        
        \item \textbf{NDVI Band Circuit:} The high entropy ($H=7.01$) and high flatness ($F=0.85$) of the NDVI band necessitated an architecture with both a wide state space and high expressivity. This is an 8-qubit, 3-layer circuit (Fig.~\ref{fig:quantum_circuits}-b)), which meets the strict $Q \gtrsim H$ constraint.
        
        \item \textbf{EVI Band Circuit:} The  circuit is most influenced by the EVI band's high amount of variability ($\sigma^2 \approx 6512$). The 4-qubit, 4-layer circuit shown in Fig.~\ref{fig:quantum_circuits}-c) is designed for maximum functional flexibility.
        
        \item \textbf{Entropy Band Circuit:} This circuit responds to the specific information profile of high entropy ($H=7.16$) and flatness ($F=0.94$) along with low edge density ($ED \approx 0.14$), having 7 qubits in order to meet the $Q \gtrsim H$ requirement as shown in Fig.~\ref{fig:quantum_circuits}-d).
    \end{itemize}
    Similarly, the circuits were designed for SAT-6 dataset by following the framework and the dataset's classical metrics as recorded in Table~\ref{tab:classical_stats}.
    \item \textbf{Feature Aggregation:} All of the outputs for all quantum circuits for that patch are concatenated into one large 37-dimensional feature vector. These vectors are reshaped into a spatial feature map of dimensions $H' \times W'$ where $H'$ and $W'$ denote the spatial grid dimensions.
\end{itemize}

\subsubsection{Classical Classifier Head:}
The aggregated quantum feature map is processed by a spatial attention head, where a $1\times1$ convolution followed by spatial softmax producing an attention map highlighting salient patches. This map is applied element-wise to the quantum features, which are then spatially summed to obtain a fixed-length context vector. The context vector is passed through a multilayer perceptron (MLP) with BatchNorm, ReLU activation, and Dropout to produce the final classification logits.

We evaluate two QMC-Net variants, with and without a residual block. The residual block, consisting of a Conv2d–BatchNorm2d–ReLU sequence, is inserted between quantum feature aggregation and the Spatial Attention Head, with its output added to the aggregated features via a skip connection.

The model is trained on an NVIDIA A100-PCIE-40GB GPU for 100 epochs using a batch size of 128 and $8\times8$ non-overlapping patches. Training employs the Adam optimizer with a learning rate of 0.0005 and categorical cross-entropy loss.

\section{Results and Analysis}
We evaluate our proposed QMC-Net on the EuroSAT and SAT-6 dataset, benchmarking its performance and parameter efficiency against a comprehensive set of state-of-the-art classical and hybrid quantum-classical models. The results, summarized in Table~\ref{tab:sat6_acc} and \ref{tab:eurosat_acc}, demonstrate the compelling advantages of our data-driven design framework.

\begin{table}[h!]
    \centering
    \setlength{\tabcolsep}{3pt}

    \begin{minipage}[t]{0.40\textwidth}
        \centering
        \caption{SAT-6 SOTA Comparison}
        \label{tab:sat6_acc}
        \vspace{2pt} 
        
        \resizebox{\linewidth}{!}{
            \renewcommand{\arraystretch}{1.25}
            \begin{tabular}{|l|c|}
                \hline
                \textbf{Source and Model} & \textbf{Acc.(\%)} \\
                \hline
                \multicolumn{2}{|l|}{\textit{State-of-the-Art Classical Baselines}} \\
                \hline
                Basu \cite{basu2015deepsat} (DBN) & 76.47 \\
                \hline
                Basu \cite{basu2015deepsat} (SDAE) & 78.43 \\
                \hline
                Basu \cite{basu2015deepsat} (CNN) & 79.10 \\
                \hline
                Basu \cite{basu2015deepsat} (DeepSAT) & 93.92 \\
                \hline
                Zhong \cite{zhong2017satcnn} (TradCNN Z-Score) & 98.34 \\
                \hline
                Zhong \cite{zhong2017satcnn} (SatCNN Linear) & 99.58 \\
                \hline
                Zhong \cite{zhong2017satcnn} (SatCNN Z-Score) & 99.61 \\
                \hline
                Liu \cite{Liu_2019} DeepSAT V2 & 99.84 \\
                \hline
                \multicolumn{2}{|l|}{\textit{Hybrid Quantum Classical Architecture}} \\
                \hline
                Sebastianelli \cite{sebastianelli2021circuit} (Ry Circ.) & 92.86 \\
                \hline
                Sebastianelli \cite{sebastianelli2021circuit} (Bellman Circ.) & 89.60 \\
                \hline
                Sebastianelli \cite{sebastianelli2021circuit} (Real Amp. Circ.) & 94.79 \\
                \hline
                \multicolumn{2}{|l|}{\textit{\textbf{Proposed Model}}} \\
                \hline
                \textbf{QMC-Net} & \textbf{99.34} \\ 
                \hline
                \textbf{QMC-Net+Residual} & \textbf{99.39} \\ 
                \hline
            \end{tabular}
        }
    \end{minipage}
    \hfill
    \begin{minipage}[t]{0.58\textwidth}
        \centering
        \caption{EuroSAT Performance Comparison}
        \label{tab:eurosat_acc}
        \vspace{2pt}
        
        \resizebox{\linewidth}{!}{
            \renewcommand{\arraystretch}{1.2}
            \begin{tabular}{|l|l|r|c|}
                \hline
                \textbf{Source} & \textbf{Model / Config.} & \textbf{\# Params} & \textbf{Acc.(\%)} \\
                \hline
                \multicolumn{4}{|l|}{\textit{State-of-the-Art Classical Baselines}} \\
                \hline
                Saricayir \cite{saricayir2025efficientnet} & EfficientNet-B0 & $\approx$5.3 M & 98.00 \\
                \hline
                Helber \cite{helber2019eurosat} & GoogleNet & $\approx$7 M & 98.00 \\
                \hline
                Li \cite{li2020deep} & ResNet-18 & $\approx$11 M & 98.00 \\
                \hline
                Helber \cite{helber2019eurosat} & ResNet-50 & $\approx$25 M & 98.00 \\
                \hline
                \multicolumn{4}{|l|}{\textit{Hybrid Quantum-Classical Baselines}} \\
                \hline
                Sebastianelli \cite{sebastianelli2021circuit} & Ry Circuit & 42k+24q & 79.00 \\
                \hline
                Sebastianelli \cite{sebastianelli2021circuit} & Bellman Circ. & 42k+24q & 84.00 \\
                \hline
                Sebastianelli \cite{sebastianelli2021circuit} & Real Amp. & 42k+48q & 92.00 \\
                \hline
                Zhang \cite{zhang2023remote} & - & 21M+72q & 96.60 \\
                \hline
                Sebastianelli \cite{sebastianelli2021circuit} & Coarse-to-Fine & 170k+96q & 97.00 \\
                \hline
                Sebastianelli \cite{sebastianelli2025quanv4eo} & Quanv4EO+Clust & 42k+16q & 96.00 \\
                \hline
                Sebastianelli \cite{sebastianelli2025quanv4eo} & Quanv4EO+Auto & 42k+16q & 91.00 \\
                \hline
                Sebastianelli \cite{sebastianelli2025quanv4eo} & Quanv4EO+DL & 42k+16q & 93.00 \\
                \hline
                \multicolumn{4}{|l|}{\textit{\textbf{Proposed Model}}} \\
                \hline
                \textbf{Ours} & \textbf{QMC-Net} & \textbf{8.9k+543q} & \textbf{93.80} \\
                \hline
                \textbf{Ours} & \textbf{QMC-Net+Residual} & \textbf{21.3k+543q} & \textbf{94.69} \\
                \hline
            \end{tabular}
        }
    \end{minipage}
\end{table}

Our hybrid quantum–classical network achieves accuracies of 99.34\% and 99.39\% on the SAT-6 dataset (Table~\ref{tab:sat6_acc}), outperforming the baseline DeepSAT framework and matching recent state-of-the-art models \cite{zhong2017satcnn, Liu_2019}. For a fair comparison, we re-implemented the hybrid architectures of Sebastianelli et al. \cite{sebastianelli2021circuit} on the same dataset, where QMC-Net consistently achieved superior performance. Notably, this accuracy is obtained with a highly compact model: the classical component without the residual block contains only 10.5k parameters, while the quantum component uses just 824 parameters. This efficiency enables effective exploitation of correlations across RGB and NIR-derived channels, highlighting the expressiveness of parameterized quantum circuits as a low-resource alternative to conventional deep learning models for remote sensing tasks.

In Table~\ref{tab:eurosat_acc} for EuroSAT dataset, at the upper end, large-scale classical models such as EfficientNet-B0 and ResNet-50 \cite{saricayir2025efficientnet, helber2019eurosat}, alongside heavyweight hybrid architectures like those from Zhang et al. \cite{zhang2023remote} and the ensemble-based "Coarse-to-fine" model \cite{sebastianelli2021circuit}, achieve state-of-the-art accuracies in the 96-98\% range. However, this performance is predicated on massive classical backbones, with parameter counts ranging from 5 million to over 21 million. These models represent a "classically-dominated" paradigm.

The primary contribution of our work is demonstrated by comparing QMC-Net to the more "quantum-centric" models from Sebastianelli et al. \cite{sebastianelli2021circuit}. Our model achieves a test accuracy of 93.80\% and 94.69\%, substantially outperforming their best single-circuit model (Real Amplitudes at 92.00\%). Crucially, this is achieved with a classical backbone that is nearly five times smaller (8.9k vs. 42.3k parameters). This indicates that our data-driven framework allows the bespoke quantum components to perform more effective and meaningful feature extraction, thus reducing the burden on the classical part of the network.

To validate our core hypothesis that data-tailored circuits outperform monolithic designs, we conduct an architectural ablation study comparing QMC-Net with four variants: a classical-only model and three monolithic hybrid models using a single quantum ansatz across all channels. As shown in Table~\ref{tab:ablation_arch}, the classical-only baseline achieves only 84.54\% accuracy, confirming the significant contribution of quantum feature extractors. Among monolithic hybrids, increased entanglement alone does not ensure better performance, as the Bellman ansatz underperforms the simpler Ry circuit. The generic Real Amplitudes ansatz performs best among monolithic designs, reaching 92.04\% accuracy, yet remains inferior to the proposed band-specific approach.

\begin{table}[!h]
    \caption{Architectural Ablation: Band-Specific vs. Monolithic Ansatze.}
    \label{tab:ablation_arch}
    \centering
    \scriptsize
    \renewcommand{\arraystretch}{1.4}
    \setlength{\tabcolsep}{6pt}
    \begin{tabular}{|l|c|c|c|c|}
        \hline
        \textbf{Ansatz for All Bands} & \textbf{Accuracy (\%)} & \textbf{Precision} & \textbf{Recall} & \textbf{F1} \\
        \hline
        No Quantum Circuits in QMC-Net & 84.54 & 84.40 & 84.50 & 84.30 \\
        \hline
        Ry Circuit & 91.87 & 92.10 & 91.90 & 91.80 \\
        Bellman Circuit & 90.98 & 91.30 & 91.00 & 90.90 \\
        Real Amplitudes & 92.04 & 92.00 & 92.00 & 92.00 \\
        \hline
        \textbf{Band-Specific (Ours)} & \textbf{93.80} & \textbf{93.80} & \textbf{93.80} & \textbf{93.80} \\ 
        \hline
    \end{tabular}
\end{table}

The key result is the clear gain achieved by moving from a monolithic to a tailored design. As shown in Table~\ref{tab:eurosat_acc}, models using a \textit{single} Ry or Bellman circuit achieve only 79\% and 84\% accuracy, while this monolithic variants in our QMC-Net already reach $\approx$91-92\% accuracy, reflecting the strength of the overall architecture. Importantly, QMC-Net further improves performance to 93.80\% by employing band-specific circuits, demonstrating that data-driven circuit tailoring provides an additional and meaningful performance boost.

\begin{table}[!h]
    \caption{Additive Ablation Study on Band Contribution.}
    \label{tab:ablation_band}
    \centering
    \scriptsize
    \renewcommand{\arraystretch}{1.4}
    \setlength{\tabcolsep}{6pt}
    \begin{tabular}{|c|c|c|c|c|c|c|}
        \hline
        \multicolumn{6}{|c|}{\textbf{Bands Included}} & \textbf{Accuracy (\%)} \\
        \hline
        \textbf{R} & \textbf{G} & \textbf{B} & \textbf{EVI} & \textbf{NDVI} & \textbf{Entropy} & \textbf{Result} \\
        \hline
        \checkmark & \checkmark & \checkmark & & & & 80.86 \\
        \hline
        \checkmark & \checkmark & \checkmark & \checkmark & & & 88.41 \\
        \checkmark & \checkmark & \checkmark & & \checkmark & & 90.88 \\
        \checkmark & \checkmark & \checkmark & & & \checkmark & 85.65 \\
        \hline
        \checkmark & \checkmark & \checkmark & \checkmark & \checkmark & & 90.39 \\
        \checkmark & \checkmark & \checkmark & \checkmark & & \checkmark & 89.48 \\
        \checkmark & \checkmark & \checkmark & & \checkmark & \checkmark & 91.95 \\
        \hline
        & & & \checkmark & \checkmark & \checkmark & 87.53 \\
        \hline
        \textbf{\checkmark} & \textbf{\checkmark} & \textbf{\checkmark} & \textbf{\checkmark} & \textbf{\checkmark} & \textbf{\checkmark} & \textbf{93.80} \\
        \hline
    \end{tabular}
\end{table}

To assess the contribution of each engineered data channel and its corresponding circuit, we perform an additive ablation study starting from an RGB-only baseline and progressively incorporating biophysical and textural features. As shown in Table~\ref{tab:ablation_band}, the RGB-only model achieves 80.86\% accuracy, while adding a single biophysical index yields a substantial improvement, with NDVI (+10.02\%) providing a larger gain than EVI (+7.55\%), validating the effectiveness of the proposed feature design.

The study also highlights non-trivial feature interactions. While combining RGB with both EVI and NDVI leads to a strong result (90.39\%), the combination of `RGB + NDVI + Entropy` achieves an even higher accuracy of 91.95\%, suggesting that the textural information from the Entropy band offers a complementary feature set to the NDVI that is more discriminative than the information provided by EVI in this context. This shows that each channel, processed by a tailored quantum circuit, contributes complementary information, and that jointly exploiting spectral, biophysical, and textural cues is key to achieving state-of-the-art performance.

\section{Conclusions, Discussions, and Future Work}
In this work, we re-examine prevailing practices in hybrid quantum model design for multi-band remote sensing and argue against the use of generic, fixed ansatzes across heterogeneous spectral channels. Instead, we propose a data-driven workflow that explicitly links the statistical complexity of each channel to appropriate quantum circuit hyperparameters. This establishes a principled pathway from data characteristics to band-aware quantum architectures, moving beyond ad hoc and uniform circuit choices.

\noindent Building on this framework, we introduce \textbf{QMC-Net}, which processes six-channel inputs derived from multi-spectral remote sensing data and constructs band-specific quantum circuits accordingly. \textbf{QMC-Net} achieves \textbf{93.80\%} accuracy on \textbf{EuroSAT} and \textbf{99.34\%} on \textbf{SAT-6}, while the residual-enhanced variant further improves performance to \textbf{94.69\%} and \textbf{99.39\%}, respectively. Extensive ablation studies support our hypothesis that tailoring quantum circuit architectures to data complexity yields a more efficient and effective design than monolithic hybrid models under NISQ constraints.

These findings suggest several directions for future work. The framework can be extended to dense prediction tasks, such as remote sensing image segmentation, and integrated with quantum self-supervised or weakly supervised learning to better leverage large-scale unlabeled data. It also naturally supports neural architecture search (NAS) for automated optimization of quantum circuits under NISQ constraints and can be generalized to additional remote sensing modalities, including hyperspectral imagery, SAR, and multi-sensor settings.

In summary, this work contributes both a \textbf{high-performing hybrid quantum--classical model} and a \textbf{theoretically grounded methodology} for designing data-aware quantum circuits, providing a principled foundation for scalable and efficient hybrid quantum systems in remote sensing and beyond. This positions data-aware quantum circuit design as a viable paradigm for scalable quantum-enhanced perception.

\bibliography{references}

\newpage
\appendix

\setcounter{section}{0}
\setcounter{figure}{0}
\setcounter{table}{0}
\renewcommand{\thefigure}{S\arabic{figure}}
\renewcommand{\thetable}{S\arabic{table}}

\begin{center}
    {\LARGE \bfseries Supplementary Material}
\end{center}

\section{Band-Specific Quantum Circuit Design for the SAT-6 Dataset}
After performing statistical analysis on the Sat-6 dataset, for the high-complexity bands, namely Red, Green, Blue, NDVI and Entropy, we discovered that these bands had consistently high Shannon Entropy values ranging from 7.08 to 7.23, with variance exceeding 1600. These statistical metrics suggest that a very rich and high-contrast feature set exists, one that demands significant representational capacity. Therefore, we developed a Complex Band Circuit comprised of 8 qubits and a circuit depth of 4 layers. To avoid the potential loss of information caused by the significant variances in the input features, we employed a Data Re-uploading technique and re-encoded the input features at each circuit layer. In addition, the Edge Density for these bands was measured at approximately 0.17 to 0.28, indicating moderate structural complexity within the bands. The spectral flatness was above 0.85, demonstrating a large area of uniformity within the bands. Therefore, we employed a Ring Entanglement topology in order to create structural correlations and break spectral symmetries by utilizing $CRZ$ gates for the Ring Entanglement topology and $PSWAP$ gates for the global mixing of all features.

\begin{figure}[!h]
    \centering
    \includegraphics[width=\textwidth]{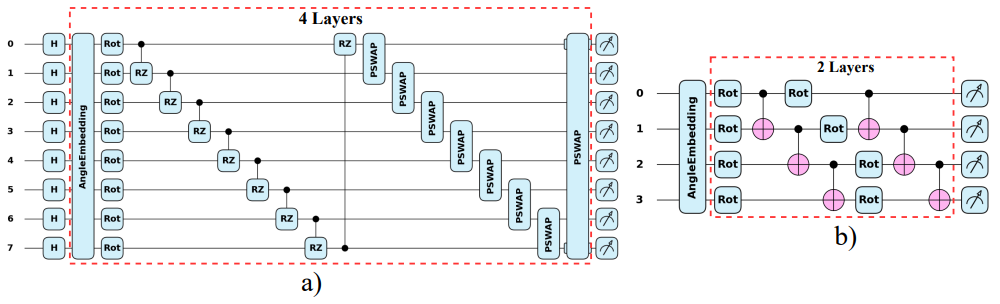} 
    \caption{Band-specific quantum circuits for SAT-6 dataset, a) For RGB, NDVI, \& Entropy bands, b) For EVI band}
    \label{fig:sat_circuits}
\end{figure}

The Enhanced Vegetation Index (EVI) Band provided a very dissimilar and distinct set of statistics when compared to the RGB, NDVI and Entropy bands, necessitating lightweight optimization. EVI had considerably lower Entropy of 4.44 and Variance of 694.8, indicating relatively less randomness and more tightly clustered data as compared to other bands. Assigning a high-capacity circuit to EVI would have resulted in wasted computational resources and overfitting. With this in mind, the EVI channel was processed using the Simple Band Circuit limited to 4 qubits and shallow depth of 2 layers with only one step for data encoding. The Edge Density for EVI was also very low, measuring at 0.0495, which means the majority of EVI consists of smooth continuous areas, thus warranting the use of a simple linear entanglement topology built with standard CNOT gates. Complex closed-loop entanglement was not required due to the low spatial complexity of the EVI band. This physics-aware resource allocation ensures that quantum computational effort is directed only toward bands with high data complexity.

\section{QMC-Net With Residual Block and Skip Connection}

\begin{figure}[!t]
    \centering
    \includegraphics[width=\textwidth]{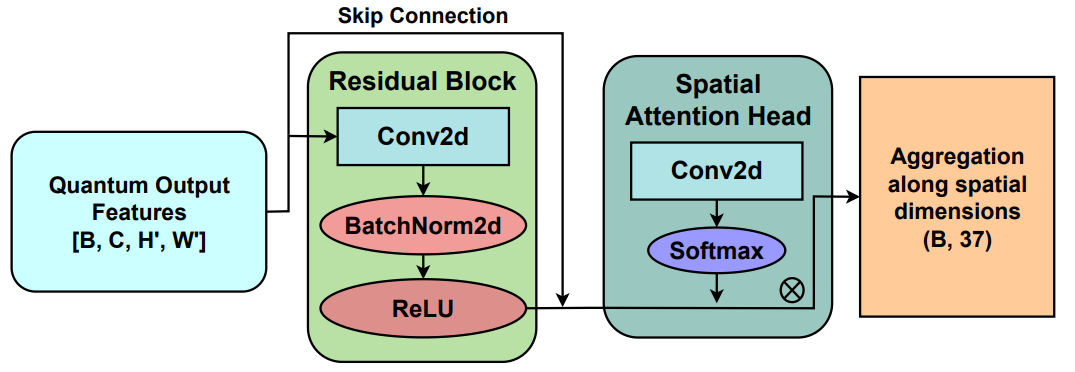} 
    \caption{Placement of Residual block in our QMC-Net Architecture}
    \label{fig:qmc_res}
\end{figure}

In order to mitigate spatial discretisation artefacts created by independent patch processing within the quantum layer, we implemented an additional classical residual refinement block immediately before the spatial attention head. This classical residual refinement block consists of a $3\times3$ convolutional layer with batch normalisation and a residual skip connection as formally defined below:
\begin{equation}
    x_{ref} = \text{ReLU}(x_{q} + \text{BN}(\text{Conv}(x_{q})))
\end{equation}
The architecture is structurally designed to provide local receptive field overlap, allowing a smooth transition between adjacent quantum patches. The residual block allows for feature interaction across patch boundaries, leaving the Spatial Attention to compute the significance of each spatially coherent structure as opposed to disconnected patch features.

The architectural refinement provided a meritorious increase in performance; specifically, the classification accuracy of the EuroSAT dataset improved from 93.80\% to 94.69\% with this architectural change. Although the additional parameters resulted in a total number of approximately 21.8k, the model itself is very compact with regard to the number of parameters (less than 0.2\%) compared to a typical ResNet-18 architecture. This finding has shown that a minimalistic incorporation of classical spatial correlation into the quantum latent space allows for the removal of ambiguity from the quantum latent space (for example, resolving the difference between Rivers and Highways) and results in an ideal balance between computation complexity and classification accuracy.

\begin{figure}[!ht]
    \centering
    \includegraphics[width=\textwidth]{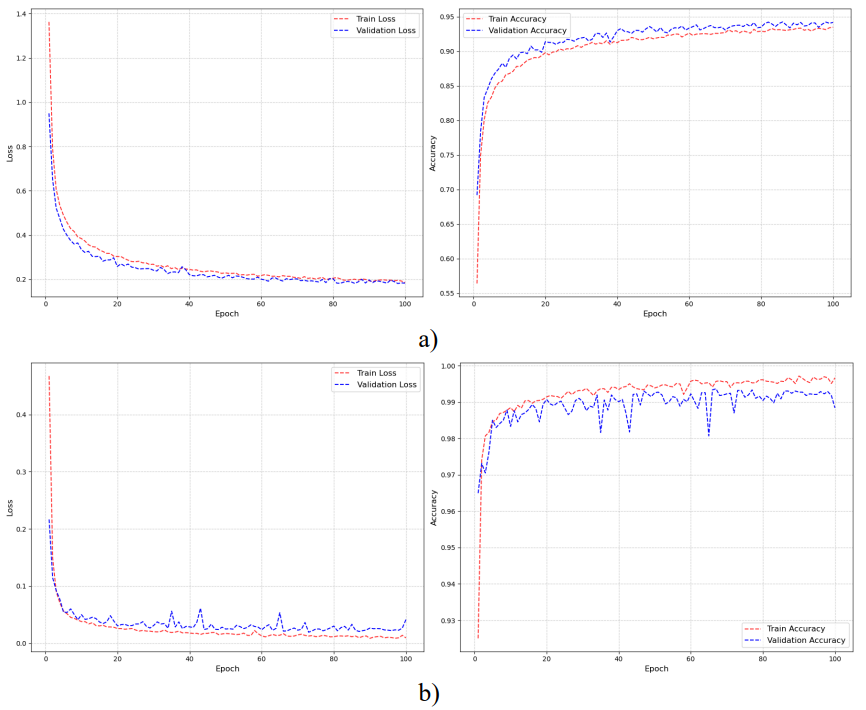} 
    \caption{Accuracy and Loss curves on a) EuroSAT dataset \& b) SAT-6 dataset}
    \label{fig:acc_loss}
\end{figure}

\section{Training Dynamics and Convergence Analysis}
With EuroSAT dataset, both training and validation losses (Fig.~\ref{fig:acc_loss}a)) steadily decline and approach one another during optimization, while model accuracy rises quickly, eventually plateauing at a relatively high level of performance demonstrating effective optimization and absence of overfitting.

For SAT-6 dataset, the final convergence point of our model's training curve occurs at a lower loss value (Fig.~\ref{fig:acc_loss}b)). The alignment between the training and validation curves indicates strong generalization properties of the model and serves to indicate its relative strength on data sets with a greater amount of class separability.

\section{Feature Space Visualization via t-SNE}
\begin{figure}[!ht]
    \centering
    \includegraphics[width=0.85\textwidth]{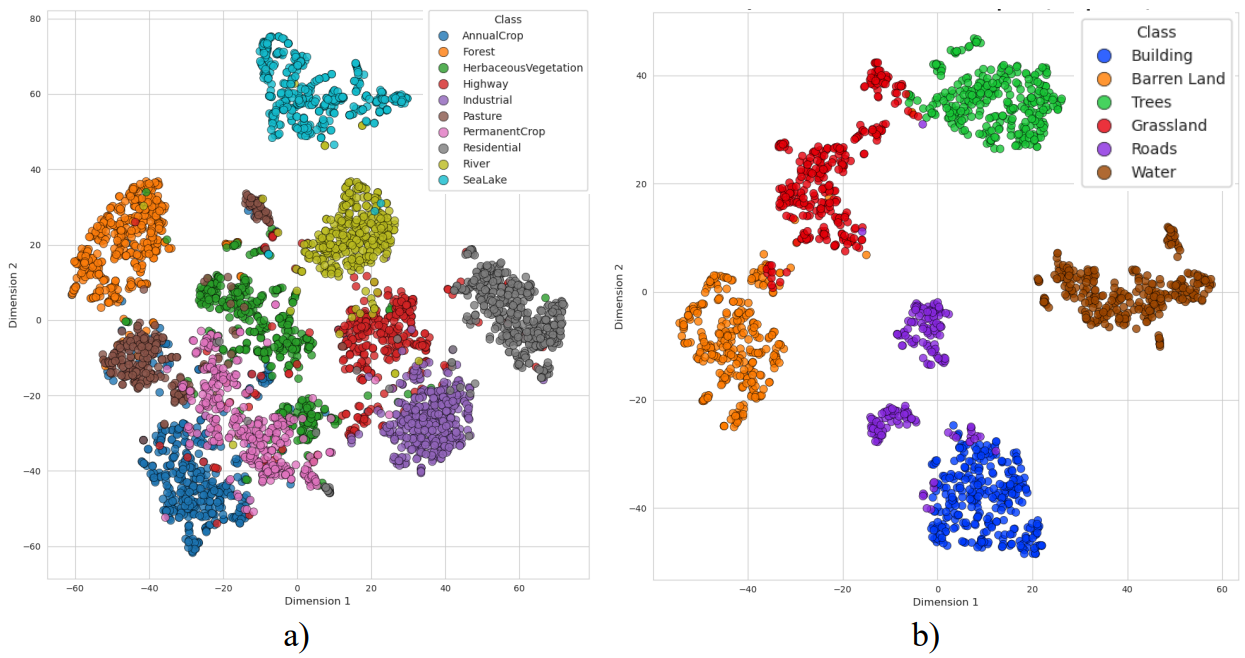} 
    \caption{t-SNE 2D plot of our trained model on a) EuroSAT dataset \& b) SAT-6 dataset}
    \label{fig:tsne}
\end{figure}

An analysis of the embedding vectors associated with the different classes of land-cover in the EuroSAT dataset reveals clearly separable clusters across different land-cover classes, as shown in Fig.~\ref{fig:tsne}a. Although mild overlap occurs in some semantically similar classes (e.g., various vegetation types), it is clear that our model has learned to extract spectral/spatial features that effectively discriminate between different classes of land.

Fig.~\ref{fig:tsne}b shows that the feature vectors for SAT-6 form well-defined boundaries between Building, Barren Land, Tree, Grassland, Road, and Water classes. These results strengthen the argument for a more compact class structure within the SAT-6 group and demonstrate the model's ability to learn class-specific representations across multiple remote sensing datasets.

\section{Confusion Matrix Analysis}

\begin{figure}[!h]
    \centering
    \begin{minipage}{0.48\textwidth}
        \centering
        \includegraphics[width=\textwidth]{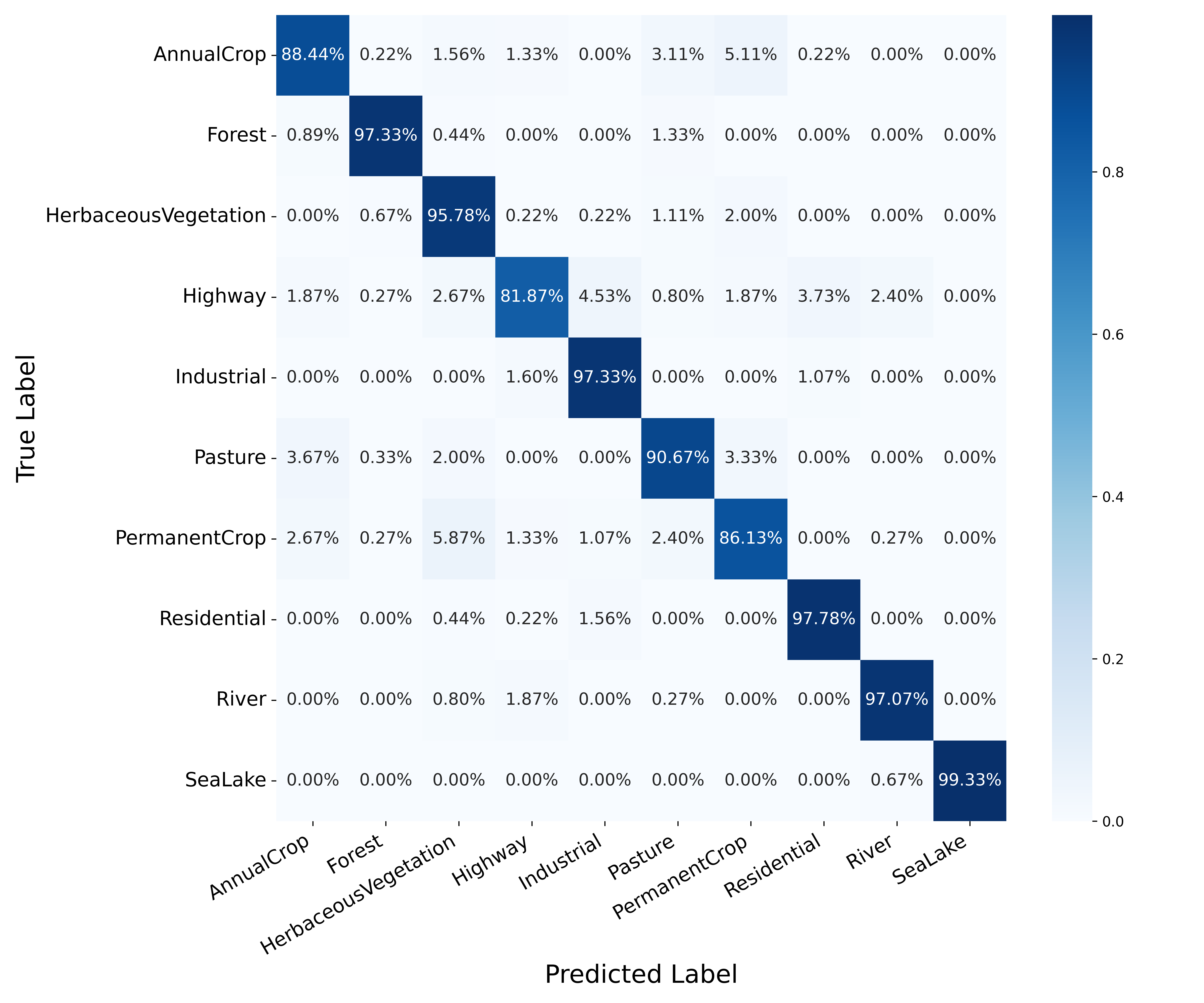}
        \label{fig:cm_eurosat}
        \\ \small (a) EuroSAT
    \end{minipage}
    \hfill
    \begin{minipage}{0.48\textwidth}
        \centering
        \includegraphics[width=\textwidth]{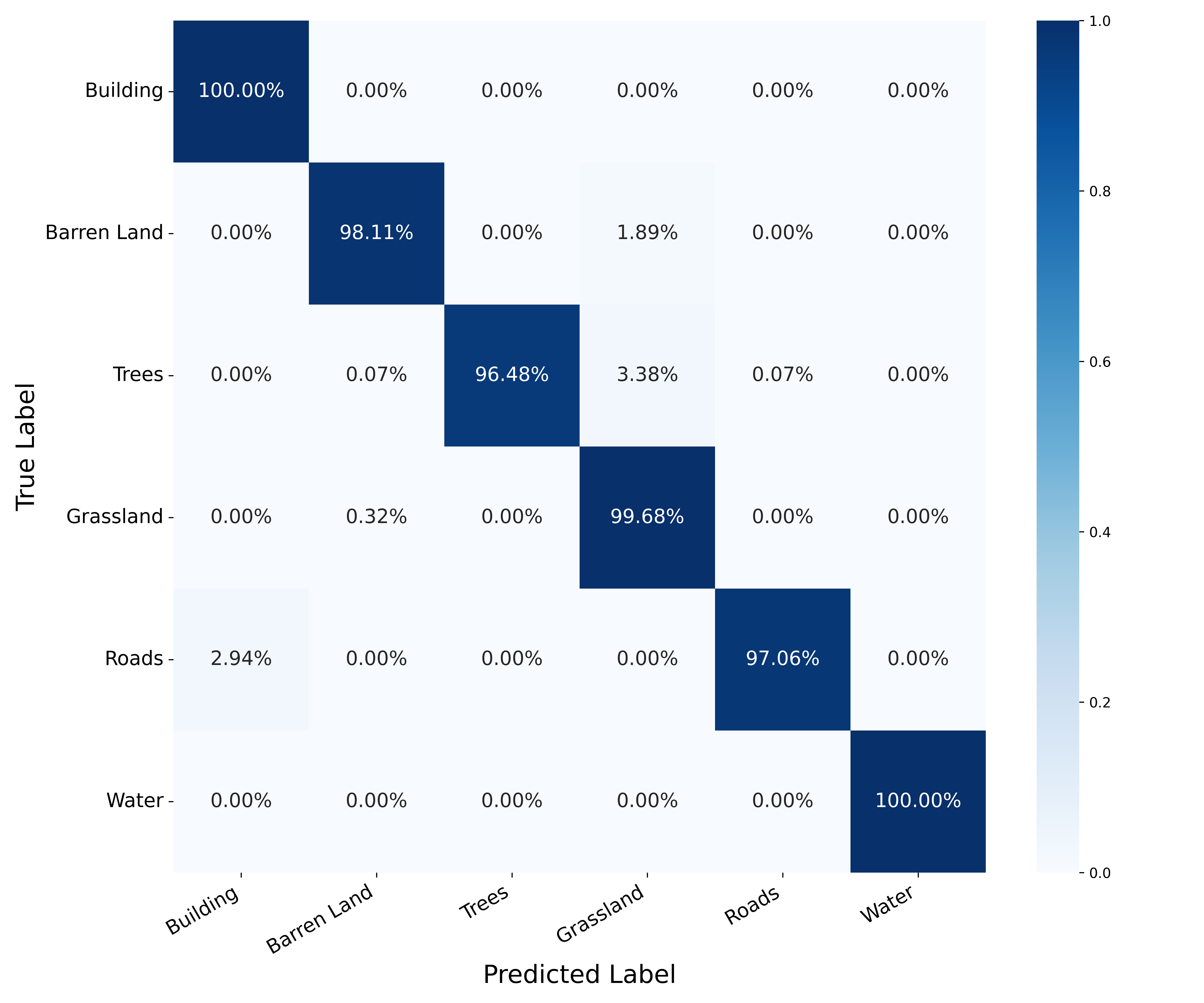}
        \label{fig:cm_sat6}
        \\ \small (b) SAT-6
    \end{minipage}
    
    \caption{Confusion matrices for (a) EuroSAT and (b) SAT-6 datasets.}
    \label{fig:confusion_matrices}
\end{figure}

A strong diagonal dominance in the confusion matrix for EuroSAT (Fig.~\ref{fig:cm_eurosat}) displays a high per-class accuracy for the majority of the land-cover types present in the dataset. The classes Forest, Industrial, Residential, River and SeaLake all show very high levels of recognition (close to 100\%), indicating that these classes maintain highly distinct spectral–spatial signatures. Confusion occurs between classes that have visually similar characteristics (for instance, AnnualCrop, PermanentCrop, Pasture and HerbaceousVegetation) as well as between linear structures such as Highway \& nearby Urbanities (e.g., Residential, etc.). A summary of these findings reveals that our model was capable of learning to represent images from the full range of lands from this study to produce highly distinctive representations while at the same time providing robust support to mitigate difficulties concerning the confusion of classes within complex remote sensing situations.


The confusion matrix for SAT-6 (Fig.~\ref{fig:cm_sat6}) exhibits an almost perfectly diagonal structure that indicates a high degree of separation between classes. Classifications of Building and Water are classified at 100\% accuracy rate. Classes Grassland, Barren Land, Trees, and Roads also have a high rate of classification accuracy, with only a minor amount of confusion occurring among them. Minor misclassifications occur due to spectral similarity between certain natural classes, although their impact is negligible. Overall, the results indicate that our model has a high degree of robustness and generalization ability on the SAT-6 dataset.


\end{document}